\documentstyle[12pt]{article}
\begin{document}
%\begin{titlepage}

\newcommand{\newc}{\newcommand}
\newc{\ra}{\rightarrow}
\newc{\lra}{\leftrightarrow}
\newc{\beq}{\begin{equation}}
\newc{\eeq}{\end{equation}}
\newc{\barr}{\begin{eqnarray}}
\newc{\earr}{\end{eqnarray}}

\def\mue{(\mu ^-,e^-)}
\def\beq{\begin{equation}}
\def\eeq{\end{equation}}
\def\al{\alpha}
\def\be{\beta}
\def\la{\lambda}
\def\ka{\kappa}
\def\del{\delta}

\begin{titlepage}
\begin{center}
\vspace*{10mm}
\vspace{10mm}

{\Large \bf The new limits of the neutrinoless $\mue$ conversion 
branching ratio }\\

\bigskip
\vspace{10mm}
T. S. KOSMAS, {\footnote{Permanent address: 
Theoretical Physics Section, University of Ioannina, GR 451 10, Greece}}
AMAND  FAESSLER

\bigskip

Institut f\"ur Theoretische Physik der Universit\"at T\"ubingen, D-72076, 
Germany.

\vspace{5mm}

and

\vspace{5mm}

J. D. VERGADOS

\bigskip

{ \it Theoretical Physics Division, University of Ioannina,
GR-451 10 Ioannina, Greece.}

\end{center}
%\maketitle

\vspace{10mm}
\begin{abstract}
{\it The nuclear physics dependence of the exotic $\mue$ conversion 
branching ratio $R_{\mu e^-}$ for the experimentally most interesting 
nuclei $^{208}Pb$ and $^{48}Ti$, is investigated in various nuclear models.
The results thus obtained are combined with the new experimental limits 
extracted at PSI for these nuclei to put bounds on the elementary particle
parameters entering $R_{\mu e^-}$ such as intermediate neutrino masses
and mixing angles as well as relevant parameters of intermediate 
supersymmetric particles (masses and mixing of s-fermions and neutralinos).}
\end{abstract}

\end{titlepage}
%%%%%%%%%%%%%%%%%%%%%%%%%%%%%%%%%%%%%%%%%%%%%%%%%%%%%%%%%%%%%%%%%%%%%%%%%
\newpage
%%%%%%%%%%%%%%%%%%%%%%%%%%%%%%%%%%%%%%%%%%%%%%%%%%%%%%%%%%%%%%%%%%

%\centerline{\large \bf 1. Introduction}
%\bigskip

There is a plethora of processes predicted by many extensions of the 
standard model like  grand unification, supersymmetric theories etc., 
which violate the lepton family quantum numbers \cite{Walter}-\cite{Depom}. 
Among them the neutrinoless conversion of a bound muon to an electron,
\beq
(A,Z)\, + \,\mu^-_b \, \ra \, e^-\, + \,(A,Z)^*
\label{eq:1.1}
\eeq
which  violates the muon and electron numbers, stands out as one of the 
most prominent processes to search for lepton flavour non-conservation if 
it exists \cite{Dohmen}-\cite{Bader}. The recent experiments at PSI 
\cite{Dohmen,Honec} and TRIUMF \cite{Ahmad}, have up to now put only upper 
limit on the branching ratio $R_{\mu e^-} \,=\, 
\Gamma(\mu ^-,e^-)/\Gamma(\mu^-,\nu_{\mu})$ in the values
$$
R^{Ti}_{\mu e^-} < 4.3\times 10^{-12},\quad for \,\,\, ^{48}Ti 
\,\,\, target \quad \cite{Dohmen} 
$$
$$
R^{Pb}_{\mu e^-} \,<\, 4.6\times 10^{-11}, \quad for \,\,\,^{208}Pb
\,\,\, target \quad \cite{Honec}
$$
We mention that, the limit set this year at PSI by using $^{208}Pb$ as 
target \cite{Honec}, improved over the previous limit \cite{Ahmad} by an 
order of magnitude. By progressively improving the experimental sensitivity, 
it is expected in the next few years these limits to be pushed down by two to 
three orders of magnitude \cite{Dohmen}-\cite{Bader}. 

On the theoretical side, up to now the elementary particle physics aspects of 
the $\mue$ conversion amplitudes have been investigated in the context of 
several models allowing lepton flavour violation 
\cite{Walter}-\cite{Depom,KLV89,KLV94}. On the other hand, the nuclear form 
factors, which describe the rates of process (1) have been computed employing 
various nuclear models \cite{KV90}-\cite{Berna}. In recent years some 
refinements of the quasi-particle RPA approximation, which have been popular 
in evaluating successfully the nuclear transition matrix elements entering 
processes like double beta decay (see e.g. ref. \cite{Suho}), double charge 
exchange \cite{Faess} etc, have also been employed in the investigation of 
$\mue$ conversion in the presence of nuclei \cite{KVCF}.

In the present work, the p-p and n-n quasi-particle RPA is used in evaluating 
the nuclear matrix elements entering $R_{\mu e^-}$ for the targets 
$^{208}Pb$ and $^{48}Ti$, for various mechanisms leading to reaction (1) 
utilizing the method developed in ref. \cite{KVCF}. Our goal is, by combining 
the nuclear calculations with the experimental limits quoted above, to set 
constraints on the elementary particle parameters entering various models 
involving intermediate massive neutrinos or other exotic supersymmetric 
particles (s-leptons and neutralinos) as it will be discussed below. We perform
explicit calculations of the nuclear matrix elements of Pb and Ti in the 
appropriate values of the momentum transfer by taking into account the muon 
binding energy entering both the phase space and the nuclear transition form 
factors. Furthermore, we calculate the contributions coming from Z-exchange 
diagrams which have not been included in previous studies \cite{KV90,KVCF}. 
We have paid special attention to the nucleus $^{208}Pb$, 
not only because it is currently used as target at PSI 
but, as we have previously shown \cite{Chia}, the rate of the $\mue$ 
conversion appears to attain a maximum in this region. We mention that, 
since $^{208}Pb$ is a doubly closed shell nucleus, it needs a special 
treatment in the context of quasi-particle RPA \cite{Suho}.

%\bigskip
%\centerline{\large \bf 2. The effective interaction Hamiltonian for $\mue$ 
%conversion}
%\bigskip

The nuclear calculations performed in this work rely on the effective 
interaction $\mue$ conversion Hamiltonian constructed in the framework of 
common extensions of the standard model involving mixing of intermediate 
neutrinos - gauge bosons or neutralinos - s-leptons (supersymmetry). Below we 
describe in brief the essential formalism required for our study.

\medskip
{\it 1. Intermediate neutrinos - gauge bosons.}
\medskip

It is well known that, within the minimal standard model, the strengths of the 
flavour changing interactions are related to the neutrino masses. In this 
spirit, one writes the weak neutrino eigenstates $\nu_e, \, \nu_\mu, ...$ in 
terms of the mass eigenstates $\nu_j$ (light neutrinos) and $N_j$ (heavy 
neutrinos) as
\beq
\nu_e =\sum_{j=1}^{n} U^{(1)}_{ej} \nu_j+\sum_{j=1}^{n}U^{(2)}_{ej} N_j
\label{eq:2.1}
\eeq
\beq
\nu_{\mu} =\sum_{j=1}^{n} U^{(1)}_{\mu j} \nu_j+                                           \sum_{j=1}^{n}U^{(2)}_{\mu j} N_j
\label{eq:2.2}
\eeq
where n=number of generations and $U^{(1)}_{ej}$, $U^{(1)}_{\mu j}$ are the 
elements of the charged lepton current mixing matrix for light neutrinos 
$\nu_j$ associated with the electron and the muon (see figs. 1,2). 
The corresponding quantities with superscript (2) refer to heavy neutrinos 
$N_j$.

In order to deduce the effective interaction Hamiltonian of the $\mue$ process 
at the nuclear level needed for our study, one starts from the weak vector and 
axial vector quark currents, 
\beq
J_{\lambda} \,=\, \sum_j {\bar q}_j \gamma_{\lambda}
(1 - \gamma_{5}) q_j, \qquad
J^{\pm}_{\lambda} \,=\, \sum_j{\bar q}_j \gamma_{\lambda} 
(1 - \gamma_{5}) \tau^{\pm} q^{\prime}_j
\label{eq:2.3}
\eeq
($J_{\lambda}$ neutral, $J^{\pm}_{\lambda}$ charged currents) where 
$q_j, q_j^{\prime} =u, d$ quarks (we neglect strange quark contribution into the
nucleon and consider only left-handed currents).

The next step is, to write at the nucleon level the hadronic current for each 
specific mechanism by taking the matrix elements of eq. (\ref{eq:2.3}) using
the appropriate quark model wave function for the nucleon. In this work we 
assumed a non-relativistic nucleon wave function and studied the following 
types of mechanisms:

{\it i) Photonic diagrams}: The $\mue$ conversion can proceed via the diagrams 
of figs. 1(a) and 2(a), i.e. by exchange of a virtual photon, provided the 
upper vertex is allowed. The hadronic vertex is the usual electromagnetic 
coupling. At the nucleon level it can be written as 
\beq
J^{(1)}_{\lambda} = {\bar N}_p \gamma_{\lambda} N_p 
 \,= \,{\bar N} \gamma_{\lambda} \frac{1}{2}\,(1 + \tau_3)\, N 
\qquad (photonic )
\label{eq:2.4}
\eeq
($N_p$, $N$ represent the proton, nucleon spinors) This current involves equal 
isoscalar and isovector components.

{\it ii) Non-photonic diagrams}: There is a plethora of such diagrams. The most 
obvious are the one which is mediated by Z-particle exchange shown in figs. 
1(a),(b) and 2(a),(b), and the box diagrams involving intermediate massive
neutrinos and W-bosons as in fig. 1(c) or mediated by s-leptons and neutralinos
as in fig. 2(c). In such cases the hadronic current takes the form
\beq
J_{\lambda}^{(2)} = {\bar N}\gamma_{\lambda} \frac{1}{2}\left[
(3 + f_V \beta \tau_3) - (f_V\beta^{''}+f_A \beta^{'} \tau_3)
\gamma_5 \right ] N \qquad (non-photonic)
\label{eq:2.5}
\eeq
The parameter $\beta^{''}$ takes the value unity except in Z-exchange when
it is zero. For the box diagrams we have $\beta=\beta^{'}$. In the case of 
W-boson exchange we have $\beta=5/6$ while for intermediate s-leptons we get
$\beta =0.6$. For Z-exchange we find
\beq
\beta^{'} = \frac{3}{2sin^2\theta_W}=6.90
\label{eq:2.6}
\eeq
\beq
\beta = 3(1-\frac{1}{2sin^2\theta_W})=-3.46
\label{eq:2.7}
\eeq
In the above equations $\beta =\beta_1/\beta_0$ with $\beta_0$ $(\beta_1)$ 
being the isoscalar (isovector) quantities at the quark level. $f_V$, $f_A$ 
are the vector, axial vector static nucleon form factors ($f_A/f_V = 1.24$).

The effective Lagrangian at the nucleon level, which involves both photonic 
and non-photonic contributions can be cast in the form
\beq
{\cal M} \,=\,\frac{4\pi\alpha}{q^2} \, j^{\lambda}_{(1)} J^{(1)}_{\lambda}
\, + \, \frac{\zeta}{m^2_{\mu}} \, j^{\lambda}_{(2)} J^{(2)}_{\lambda}
\label{eq:2.8}
\eeq
where $\alpha$ is the fine structure coupling constant, $\zeta$ is given by
\beq
\zeta \,= \, \frac{ G_Fm^2_{\mu}}{\sqrt 2} \qquad \,\,\, 
(W-boson \,\,\, exchange)
\label{eq:2.9}
\eeq
and $q$ is the momentum transfer which in a good approximation is written as
\beq
q \,=\, m_\mu - \epsilon -(E_f - E_{gs})
\label{eq:2.10}
\eeq
with $\epsilon$ the muon binding energy and $E_f$ ($E_{gs}$) the energy of the 
final (ground) state of the nucleus.

In eq. (\ref{eq:2.8}) $j^{\lambda}_{(1)}$ and $j^{\lambda}_{(2)}$ represent 
the leptonic currents corresponding to each mechanism i.e.

\medskip
\noindent
{\it i) photonic mechanism:}

\beq
j^{\lambda}_{(1)} = {\bar u}(p_e) \Big[ \, (f_{M1} + \gamma_5 f_{E1}) i
\sigma^{\lambda \nu} \frac{q_{\nu}}{m_{\mu}} +
\left (f_{E0} + \gamma_5 f_{M0}\right) \gamma_{\nu}  \left (
g^{\lambda \nu} - \frac {q^{\lambda} q^{\nu}}{q^2}\right )
\Big] u(p_{\mu})
\label{eq:2.11}
\eeq

\noindent
{\it ii) Non-photonic mechanism:}

\beq
j^{\lambda}_{(2)} = {\bar u}(p_e)\gamma^{\lambda}
\, {\frac{1}{2}} \, (f_1 + f_2 \gamma_5) u(p_{\mu})
\label{eq:2.12}
\eeq
where $p_e,\,\,p_\mu$ the lepton momenta and $f_{E0}$, $f_{E1}$, $f_{M0}$, 
$f_{M1}$, $f_1$, $f_2$ parameters depending on the assumed gauge model.

In the last step, the effective interaction Hamiltonian, which converts a muon 
to an electron in the nucleus $(A,Z)$, is obtained by summing over all type of 
eq. (\ref{eq:2.8}) single nucleon contributions. Behind this summation lies 
the assumption that the A nucleons interact individually with the muon field 
(impulse approximation). 
 
\medskip
{\it 2. Intermediate neutralinos - s-leptons (supersymmetric model).}
\medskip

The supersymmetry (SUSY) associates to each particle its superpartner 
(s-particle). In this way, the fundamental particles of nature are essentially 
doubled and as a consequence several new mechanisms, some of which violate the 
lepton flavour conservation, are predicted. Some typical supersymmetric 
diagrams leading to the $\mue$ conversion are shown in fig. 2. 

In the effective Hamiltonian of eq. (\ref{eq:2.8}), the hadronic current is
given by eq. (\ref{eq:2.5}) by putting the values of $\beta$ discussed before.
For the leptonic currents, in the simplest version of this model, the 
photonic and non-photonic amplitudes are simply related \cite{KLV94}. One finds:
\beq
(4\pi\alpha)f_{M1}=-(4\pi\alpha)f_{E1}=-\frac{1}{24}f 
\label{eq:2.13}
\eeq
\beq
(4\pi\alpha)f_{E0}=-(4\pi\alpha)f_{M0}=-\frac{1}{72}f 
\label{eq:2.14}
\eeq
\beq
f_{1}  =  -f_{2}=\frac{1}{16}\beta_0 f
\label{eq:2.15}
\eeq
\beq
f  =  \alpha^{2}\frac{m^2_{\mu}}{\bar{m}^2}\tilde\eta
\label{eq:2.16}
\eeq
\beq
\tilde\eta  =  \frac{\left(\delta m^2_{ll}\right)_{12}}{\bar{m}^2} 
\label{eq:2.17}
\eeq
\beq
\zeta  =1.0 
\label{eq:2.18}
\eeq
where $\bar m^2$ is the average of the square of the masses of the s-fermions
entering the loop and $\beta_0=5/9$. The parameter $\bar \eta$ depends  
on the details of the model \cite{KLV94}. 

The Z-exchange contribution in SUSY models like the one just discussed yields 
\cite{VK}
\beq
f_{1}  =  -f_{2}=\alpha^{2}\xi\tilde\eta\frac{m^2_{\mu}}{m^2_Z} 
\label{eq:2.19}
\eeq
where the parameter $\xi$ depends on the details of the model. It vanishes
if one ignores the Higgsino components of the neutralinos as e.g. in the case
of pure photino. It also vanishes if the  probabilities of finding the two
Higgsinos in the neutralino are equal. In the model $\#$2 of ref. \cite{JDV96} 
it takes the value $\xi =2.8\times 10^{-2}$. Clearly, the Z-exchange is more
favored in models in which $\bar{m}^2$ is larger than $m^2_Z$. 
Accurate formulas and details will be presented elsewhere \cite{VK}. 

%\bigskip
%\centerline{\large \bf 3. The $\mue$ conversion branching ratio}
%\bigskip

Experimentally, the branching ratio $R_{\mu e^-}$ is quite interesting quantity.
From a theoretical point of view, in general it is not easy to formulate an 
expression containing contributions from both mechanisms, photonic and 
non-photonic ones. For this reason, the photonic and non-photonic contributions
are usually discussed separately.

In the case of the coherent channel, however, one can separate the dependence 
on the nuclear physics from the leptonic form factors in the branching ratio 
$R_{\mu e^-}$, provided that the nuclear vertex is calculated in the Born 
approximation and the density of the muon bound state inside the nucleus is 
approximated by a mean wave function (see below). Under these assumptions
the muon capture rates can well be described by the Goulard-Primakoff 
function $f_{GP}(A,Z)$ \cite{Goul-Prim}. In general, it is impossible to
separate the nuclear structure aspects from the elementary particle 
parameters. This can, however, approximately be done in the models discussed 
above \cite{KLV94}. Thus one can write
\beq
R_{\mu e^-} = \rho \, \gamma
\label{eq:3.1}
\eeq
The function $\gamma(A,Z)$ contains all the nuclear dependence of 
$R_{\mu e^-}$ and the quantity $\rho$ is independent on nuclear physics. 
They are defined as follows:

\bigskip
\noindent
{\it i) Neutrino mixing models}:

\beq
\gamma \, = \, \frac{E_e p_e}{m_\mu^2} \,
\frac{\vert ME \vert^2}{G^2 Z f_{GP}(A,Z)}
\label{eq:3.2}
\eeq

\noindent
($G^2 \approx 6$, ${\bf p}_e = -{\bf q}$) where $|ME|$ represent the nuclear 
matrix elements given in terms of the proton ($F_Z(q^2)$) and neutron 
($F_N(q^2)$) nuclear form factors. In the special case of photonic diagrams 
$|ME|$ is proportional to the elastic or inelastic form factor 
$F_Z(q^2)$ depending on which final state is populated. Then, 
$\gamma$ takes the form
\beq
\gamma_{ph} \,=\, \frac{E_e P_e}{m_\mu^2}\, 
\frac{Z |F_Z(q^2)|^2 }{G^2 f_{GP}(A,Z)}
\label{eq:3.3}
\eeq
The quantity $\rho$ in eq. (\ref{eq:3.1}), in the case of left-handed 
currents only, takes the form

\barr
\rho &=& (4\pi\alpha)^2\frac{\vert f_{M1}+f_{E0}\vert ^2+  
         \vert f_{E1}+f_{M0}\vert^2}{(G_Fm^2_{\mu})^2}
\nonumber \\
   &=& \frac{9\alpha^2}{64\pi^2}\vert\frac{m^2_e}{m^2_W}\eta_{\nu}+\eta_N 
	 \vert ^2 
\qquad \,\, (photonic \,\, diagrams) \qquad
\label{eq:3.4}
\earr
\barr
\qquad \qquad \rho &=&\frac{\vert\beta_0f_1\vert^2+\vert\beta_0f_2\vert^2}{2}
\nonumber \\
     &=& \frac{9}{64\pi^4}(G_Fm^2_W)^2\vert 30\frac{m^2_e}{m^2_W}\eta_{\nu}
	 +\eta_N\vert ^2
\quad (non~-~photonic \,\, diagrams) 
\label{eq:3.5}
\earr
\noindent
The lepton violating parameters associated with intermediate light
$(\eta_{\nu})$ or heavy $(\eta_N)$ neutrinos depend on the gauge model
considered.
They are given by
\beq
\eta_\nu = \sum_j U^{(1)}_{ej} U^{(1)*}_{\mu j} \frac{m^2_j}{m^2_e}, \qquad
 \eta_N = \sum_j U^{(2)}_{ej} U^{(2)*}_{\mu j} \frac{m^2_W}{M^2_j}
\left( -2 ln \frac{ M^2_j}{m^2_W} +3 \right) 
\label{eq:3.6}
\eeq

\bigskip
\noindent
{\it ii) Neutralino and s-lepton mixing models}:
Flavour violation in these models occurs via the s-lepton mixing provided
that it is different from the corresponding charged lepton mixing. It also
involves intermediate neutralinos. Following \cite{KLV94} we will assume
that the  dominant contribution comes from the photino (SUSY partner of the 
photon). Under some plausible assumptions \cite{KLV89} and neglecting the   
Z-exchange we find that the function $\gamma(A,Z)$ can be written as
\beq
\gamma \,=\, \zeta \,\Big( \frac{13}{12} + \frac{1}{2} 
\,\frac{N}{Z} \, \frac{F_N}{F_Z} \Big)^2 \, \gamma_{ph}
\label {eq:3.7}
\eeq
The quantity $\rho$ in this model can be written as
\beq
\rho \,=\,\frac{1}{288}\frac{\alpha^4}{(G_F {\bar m}^2)^2}\,
|{\bar \eta}|^2 
\label{eq:3.9}
\eeq
\noindent
where ${\bar \eta}$ is given by eq. (\ref{eq:2.17}).
In the case of Z-exchange \cite{VK} we find
\beq
\gamma \,=\, 0.053 \Big(1  - 14.04\,\frac{N}{Z} \, \frac{F_N}{F_Z}
\Big)^2 \gamma_{ph}
\label {eq:3.11}
\eeq
The quantity $\rho$ is now given by
\beq
\rho \,=\,5.5\times 10^{-4}\frac{\alpha^4}{(G_F {\bar m}^2)^2}\,
|{\bar \eta}|^2 
\label{eq:3.13}
\eeq
We stress that for the coherent process, in the models discussed above,
the only variable of the elementary particle sector, which enters 
the function $\gamma$ is the parameter $\beta =\beta_1/\beta_0$. 
Once $\gamma(A,Z)$ is known, e.g. by nuclear model calculations,
from eq. (\ref{eq:3.1}) one can extract information about the 
interesting parameter $\rho$ from the experimental data and compare it 
with the value predicted for various mechanisms by the gauge models. 
Such kind of calculations we include below.
 
%%%%%%%%%%%%%%%%%%%%%%%%%%%%%%%%%%%%%%%%%%%%%%%%%%%%%%%%%%%%%%%%%%
%\bigskip
%\centerline{\large \bf 4. Results and discussion}
%\bigskip

The task of studying the nuclear physics aspects of the exotic
$\mue$ conversion process is to evaluate the function $\gamma (A,Z)$ 
of eqs. (\ref{eq:3.2}), (\ref{eq:3.3}), (\ref{eq:3.7}) and (\ref{eq:3.11}). 
In general, the nuclear matrix elements entering $\gamma(A,Z)$ 
depend on the final nuclear state populated during process (1).
What, however, is more interesting is the coherent process, since then 
the rate is free of the background from bound muon decay \cite{Bader}.
Furthermore, extensive calculations have shown that the coherent channel
dominates the $\mue$ conversion throughout the periodic table \cite{Chia}
(for the isotopes studied in the present work see table 2 below).

If we assume an average value for the muon wave function, so that it cancels 
in the branching ratio, the coherent matrix elements needed in the computation 
of $\gamma(A,Z)$ can be written as

\beq
M^2_{coh}(q^2) =   \, \left[ \, 1 + Q(\beta) \,
\frac{N}{Z} \, \frac{F_N(q^2)}{F_Z(q^2)} \right ]^2 \,
Z^2F_Z^2(q^2)
\label{eq:4.1}
\eeq

\noindent
where $Q(\beta)$, for the models discussed in the present work, takes the values

\beq
Q(\beta) = \frac{3 - f_V \beta}{3 + f_V \beta} \, = \,
\left\{ \begin{array}{ l@{\quad \quad} l}
0,    & $ $ photonic $ $ diagrams, $ $ $ $  \beta =3 \\
13/23,& $ $ W-boson $ $ exchange, $ $ $ $  \beta =5/6 \\
2/3,  & $ $ SUSY $ $ Photonic + Box $ $ diagrams, $ $ $ $  \beta = 3/5 \\ 
-14.04,  & $ $ SUSY $ $Z-exchange, $ $ $ $ \beta = -3.46 \\ 
\end{array} \right. 
\label{eq:4.2}
\eeq
 
\noindent
Notice that, in the photonic case $M^2_{coh} \,=\, Z^2F_Z^2(q^2)$ (only 
protons of the considered nucleus contribute, see eq. (\ref{eq:2.4})). 

In table 1, we show the results of the nuclear form factors and the
coherent matrix elements for $^{208}Pb$ and $^{48}Ti$, 
calculated in the context of quasi-particle RPA using the method of
ref. \cite{KVCF}.
For comparison, in this table the results of ref. \cite{KV90}
(shell model) and those of ref. \cite{Chia} (local density 
approximation, LDA) are also presented.
The quasi-particle RPA form factors are in good agreement with experimental
electron scattering data \cite{Vries} also listed in table 1.

For the benefit of the reader we provide here a brief description 
of the main steps followed in the computational procedure.
At first, we have chosen an appropriate model space: 
for $^{208}Pb$ 18 levels above the core $^{100}_{50}Sn$ and
for $^{48}Ti$ 16 levels without core 
(for the calculations of $^{48}Ti$ in ref. \cite{KVCF} 
we used a smaller model 
space, consisted of only 10 levels without core).
In order to satisfy the convergence of BCS equations in the 
case of the doubly closed shell nucleus $^{208}Pb$, we 
determined the strength parameters of pairing for protons and
neutrons from the experimental proton and neutron pairing gaps of
the $(N-2,Z+2)$ neighbor nucleus
$^{208}_{124}Po$, following the procedure used recently in the study
of double beta decay \cite{Suho}. 

We have also evaluated the incoherent rate, i.e. the matrix elements
from the initial (ground) state (a $0^+$ state) to every excited state 
$|f>$ included in the chosen model space of the studied nucleus,
by calculating explicitly the contribution of each individual channel,
for photonic and non-photonic diagrams. These results are denoted as 
$M^2_{inc}$ and are shown in table 2 separately for the vector ($M_V$) and 
axial vector ($M_A$) components of the hadronic current eq. 
(\ref{eq:2.5}) ($M^2_{inc} = M_V + M_A$). Obviously, in the 
photonic mechanism only the vector component contributes, but in the 
non-photonic one both the vector and axial vector components give non-zero 
contributions. For each isotope studied we found that, the main contribution 
to the incoherent rate comes from the low-lying excited states and that 
high-lying excited states contribute negligible amounts. Such state-by-state 
calculations of the incoherent channel for a set of nuclei throughout the 
periodic table are discussed in ref. \cite{KVCF,KFSV}.

From the coherent and total matrix elements, $M^2_{tot}=M^2_{coh}+M^2_{inc}$, 
another useful quantity of the $\mue$ conversion, the ratio 
$\eta = M^2_{coh}/ M^2_{tot}$ which expresses the portion exhausted by the 
coherent rate in the total branching ratio $R_{\mu e^-}$, can be obtained
(see table 2). We found that, for both isotopes and all mechanisms studied, 
$\eta \ge 90\%$, which implies that the coherent rate dominates the $\mue$ 
conversion process. In earlier calculations the ratio $\eta$ was estimated 
\cite{WeiFei} to be $\eta \approx$ 83 \% in Cu region and smaller in lead 
region while in ref. \cite{Chia} we found $\eta \ge 90 \%$ throughout the
periodic table. Our present results agree well with those of ref. \cite{Chia}.

We should note that, in the present study the muon binding energy $\epsilon_b$ 
in eq. (\ref{eq:2.10}) has been taken into consideration. This property affects
significantly the nuclear form factors, especially for heavy nuclei like 
$^{208}Pb$. In light nuclei $\epsilon_b$ is negligible (for $Ti$ 
$\epsilon_b =1.450$), but in heavy elements it becomes significant 
(for $Pb$ $\epsilon_b =10.450$). In addition, the factor $E_e p_e/m^2_\mu$
in eq. (\ref{eq:3.2}), which takes into account 
the phase space in the transition matrix elements, depends on
the muon binding energy. By ignoring $\epsilon_b$, this factor becomes
unity. For $^{48}Ti$ this factor is equal to 0.97 and the neglect of 
$\epsilon_b$ is a good approximation. 
For $^{208}Pb$, however, this factor is equal to
0.81 which means that, in heavy nuclei the dependence on $\epsilon_b$ 
of the phase space cannot be ignored.

The values for $\gamma(A,Z)$ calculated from the coherent QRPA matrix elements 
are shown in table 3 and compared with the results of refs. \cite{KV90,Chia}.
We mention that, the variation of $\gamma (A,Z)$ through the periodic table 
studied in ref. \cite{KV90} exhibits a strong dependence on the neutron
excess $(N-Z)$ which mainly reflects the dependence on $(N,Z)$
of the total muon capture rate.

By putting the experimental limits \cite{Dohmen,Honec} in eq. (\ref{eq:3.1}) 
we determine upper bounds on the parameter $\rho$ for photonic, Z-exchange and 
W-box diagrams both in conventional extensions of the standard model as well 
as SUSY theories (see table 3). We note that the values of $\rho$ for $^{48}Ti$
are slightly improved over those of table 2 of ref. \cite{KVCF}, but those of 
$^{208}Pb$ are appreciably smaller than those of sect. 5.1 of ref. \cite{KV90}.
This big difference is due to the fact that, in the present work we have used 
the experimental limit of ref. \cite{Honec}. This experiment at PSI improved 
on the previous limit of $^{208}Pb$ \cite{Ahmad}, which was used in ref. 
\cite{KV90}, by an order of magnitude. For both nuclei we found that, the 
limits obtained from various nuclear models do not significantly differ from 
each other (we should stress that, the different limits for $^{208}Pb$ in the 
shell model results of table 3, are due to the neglect of $\epsilon_b$ in eq. 
(\ref{eq:2.10}) when calculating the nuclear form factors as mentioned above; 
its consideration gives 
similar results to those of LDA and QRPA for both mechanisms). This implies 
that all nuclear models studied here give about same values for $\rho$.

One can use the limits of $\rho$ to parametrize the muon number violating 
quantities entering eqs. (\ref{eq:3.4}), (\ref{eq:3.5}), (\ref{eq:3.9}) and
(\ref{eq:3.13}) and also to compare them
directly to the value given from gauge models. As an example, we quote the value
of $\rho = 8.2 \times 10^{-18}$ obtained in the supersymmetric model of ref. 
\cite{KV90} discussed above. This prediction of $\rho$ is considerably 
smaller compared to the values listed in table 3, which were extracted from 
experiment. 

%\bigskip
%\centerline{\large \bf 5. Summary and conclusions}
%\bigskip

In summary, in the present letter we have studied the nuclear physics part of 
the branching ratio of the exotic $\mu^- \ra e^-$ conversion $R_{\mu e^-}$ in 
$^{208}Pb$ and $^{48}Ti$ nuclei. These two isotopes are the most interesting 
nuclear targets to search for lepton flavour violation. $^{208}Pb$ is 
currently used at PSI in the SINDRUM II experiment.

We have calculated in a reliable way the appropriate nuclear matrix elements 
entering the branching ratio $R_{\mu e^-}$ in the context of the quasi-particle 
RPA and compared them with the results given from other nuclear models. We found
that, the coherent $\mue$ rate, which is measured from experiments, dominates 
the branching ratio $R_{\mu e^-}$ for both isotopes but it is more pronounced 
in the heavy nucleus $^{208}Pb$.

From the calculations of the  nuclear part of the branching ratio $R_{\mu e^-}$,
using the new experimental limits, especially those for $^{208}Pb$, we were 
able to estimate upper limits for the elementary sector part of the 
$\mu^- \ra e^-$ rate in common extensions of the standard model, which allow 
lepton flavour violation. These limits are very useful to fix the lepton 
violating parameters and test the various gauge models. 

%%%%%%%%%%%%%%%%%%%%%%%%%%%%%%%%%%%%%%%%%%%%%%%%%%%%%%%%%%%%%%%%%%
\bigskip
\bigskip
\bigskip
%{\it Acknowledgments:}
T.S.K. acknowledges support from the DFG No FA67/19-1 project.
He would also like to thank the members of the Institute of Theoretical 
Physics of T\"ubingen University for the warm hospitality extended to him 
while spending his sabbatical there.
%%%%%%%%%%%%%%%%%%%%%%%%%%%%%%%%%%%%%%%%%%%%%%%%%%%%%%%%%%%%%%%%%%

\newpage
%%%%%%%%%%%%%%%%%%%%%%%%%%%%%%%%%%%%%%%%%%%%%%%%%%%%%%%%%%%%%%%%%%%%%%%
\noindent
{\bf Table 1.} Quasi-particle RPA results for proton, neutron nuclear form 
factors and coherent nuclear matrix elements squared for the four cases 
discussed in the text: photonic and non photonic W-boson exchange as well as 
the two SUSY cases (photonic + box and Z-exchange). For comparison the 
experimental charge form factors ($F^{exp}_Z$) \cite{Vries} and the calculations
of shell model \cite{KV90} and LDA \cite{Chia} are shown. Note that in the 
Z-exchange the factor 0.053 of eq. (\ref{eq:3.11}) has not been included.

\vskip0.3cm
\begin{center}
\begin{tabular}{lcrrcccccccrcr}
\hline
\hline
& & & & & & & & & & & \\
  & \multicolumn{2}{c}{ Shell Model } &
       \multicolumn{2}{c}{$  $ LDA $  $ } & Exper.&
       \multicolumn{6}{c}{ Quasi-particle RPA Results} \\
\hline
& & & & & & & & & & & \\
(A,Z) & $F_Z $&$ F_N $& $ $ $ F_Z $ $ $ &$ $ $ F_N $ $ $ &
$F_Z^{exp}$&$ F_Z$ & $F_N$ & $M^2_{ph}$ & $M^2_{W-ex}$ & 
$M^2_{s-lep}$ & $M^2_{Z-ex}$ \\
\hline
& & & & & & & & & & & \\
$^{48}Ti$ 
& .543 & .528 & .528 & .506 & .532 & .537 & .514 & 139.6 &  375.2 &
429.6 & 30918.0 \\
& & & & & & & & & & & \\
$^{208}Pb$ 
& .194 & .139 & .250 & .220 & .242 & .271 & .214 & 494.7 & 1405.2 &
1618.8 & 127214.1 \\
\hline
\hline
\end{tabular}
\end{center}

\vskip2.5cm

%%%%%%%%%%%%%%%%%%%%%%%%%%%%%%%%%%%%%%%%%%%%%%%%%%%%%%%%%%%%%%%%%%%%%%%
{\bf Table 2.} 
The $\mue$ conversion matrix elements (coherent, incoherent, total) and the 
ratio $\eta = M^2_{coh}/M^2_{tot}$ for photonic and W-exchange (non-photonic) 
mechanisms involving intermediate neutrino mixing. Results for the 
experimentally most interesting nuclei, $^{208}Pb$ and $^{48}Ti$, are presented.

\vskip0.8cm
\begin{center}
\begin{tabular}{llclccrlcrc}
\hline
\hline
& & & & & & & & & & \\
&  \multicolumn{5}{l}{ Photonic $\mu^- \ra e^-$ Mechanism} & 
   \multicolumn{5}{l}{W-exchange $\mu^- \ra e^-$ Mechanism } \\
\hline
& & & & & & & & & & \\
& $M^2_{coh}$ & \multicolumn{2}{c}{$M^2_{inc}$}& $M^2_{tot}$ &  
$\eta \,(\,\%\,)$ 
& $M^2_{coh}$ & \multicolumn{2}{c}{$M^2_{inc}$}& $M^2_{tot}$ &  
$\eta \,(\,\%\,)$   \\
\hline
& & & & & & & & & & \\
(A,Z) &  & $M_V$ & $M_A$ &  &  &  & $M_V$ & $M_A$ &  & \\
\hline
& & & & & & & & & & \\
$^{48}Ti$& 139.6& 14.1& -& 153.7& 90.8&  375.2& 10.7& 2.6&  388.5& 96.6\\
& & & & & & & & & & \\
$^{208}Pb$&494.7& 14.0& -& 508.7& 97.2& 1405.2& 19.0& 6.8& 1431.0& 98.2\\
\hline
\hline
\end{tabular}
\vspace{7mm}
\end{center}
%%%%%%%%%%%%%%%%%%%%%%%%%%%%%%%%%%%%%%%%%%%%%%%%%%%%%%%%%%%%%%%%%%%%

\newpage
%%%%%%%%%%%%%%%%%%%%%%%%%%%%%%%%%%%%%%%%%%%%%%%%%%%%%%%%%%%%%%%%%%%%%%%
{\bf Table 3.} The new limits on the elementary sector part of the exotic 
$\mu-e$ conversion branching ratio extracted by using eq. (\ref{eq:3.1}) and
the recent experimental data for the nuclear targets $^{208}Pb, \,\,\,^{48}Ti$,
\cite{Dohmen,Honec}. The nuclear part of the branching ratio, described by the 
function $\gamma (A,Z)$ of eqs. (\ref{eq:3.2}), (\ref{eq:3.3}), 
(\ref{eq:3.7}) and (\ref{eq:3.11}), is also shown.

\vskip0.8cm
\begin{center}
\begin{tabular}{llrcrc}
\hline
\hline
& & & & & \\
& & \multicolumn{2}{c}{\large \bf $^{48}Ti$}&
    \multicolumn{2}{c}{\large \bf $^{208}Pb$} \\
\hline
& & & & & \\
Method & Mechanism & $\gamma(A,Z)$ & $\rho$ & $\gamma(A,Z)$ & $\rho$ \\
\hline
& & & & & \\
    & Photonic        &  9.42 & $\le$ 4.6 $\times 10^{-13}$ 
		      & 17.33 & $\le$ 2.7 $\times 10^{-12}$ \\
QRPA&W-boson exchange & 25.31 & $\le$ 1.7 $\times 10^{-13}$ 
		      & 49.22 & $\le$ 0.9 $\times 10^{-12}$ \\
    & SUSY s-leptons  & 25.61 & $\le$ 1.7 $\times 10^{-13}$ 
                      & 49.50 & $\le$ 0.9 $\times 10^{-12}$ \\
    & SUSY Z-exchange &110.57 & $\le$ 0.4 $\times 10^{-13}$ 
                      &236.19 & $\le$ 0.2 $\times 10^{-12}$ \\
\hline
& & & & & \\
LDA & Photonic        &  9.99 & $\le$ 4.3 $\times 10^{-13}$ 
         	      & 17.84 & $\le$ 2.6 $\times 10^{-12}$ \\
    &W-boson exchange & 26.60 & $\le$ 1.6 $\times 10^{-13}$ 
		      & 55.53 & $\le$ 0.8 $\times 10^{-12}$ \\
\hline
& & & & & \\
SM & Photonic         &  9.74 & $\le$ 4.4 $\times 10^{-13}$ 
                      & 10.42 & $\le$ 4.4 $\times 10^{-12}$ \\
    &W-boson exchange & 26.50 & $\le$ 1.6 $\times 10^{-13}$ 
                      & 27.43 & $\le$ 1.7 $\times 10^{-12}$ \\
\hline
\hline
\end{tabular}
\vspace{7mm}
\end{center}

%\newpage
%%%%%%%%%%%%%%%%%%%%%%%%%%%%%%%%%%%%%%%%%%%%%%%%%%%%%%%%%%%%%%%%%%%%%

\vspace{25mm}

\centerline{\large \bf Figure Captions}

\vspace{.4cm}

\noindent
{\bf Figure 1.} Typical diagrams entering the neutrinoless $\mue$ conversion:
photonic 1(a), Z-exchange 1(a),(b) and W-boson exchange 1(c), for the specific
mechanism involving intermediate neutrinos.

\vspace{25mm}

\noindent
{\bf Figure 2.} SUSY diagrams leading to the $\mue$ conversion: photonic 2(a), 
Z-exchange 2(a),(b) and box diagrams 2(c), in a supersymmetric model with 
charged s-lepton and neutralino mixing are shown. Note that, the Z-exchange in 
2(b) as well as in fig. 1(b), comes out of electrically neutral particles 
(photon exchange does not occur in these diagrams). These Z-exchange diagrams 
may be important (those of figs. 1(a) and 2(a) are suppressed by 
$m^2_{\mu}/m^2_Z$).

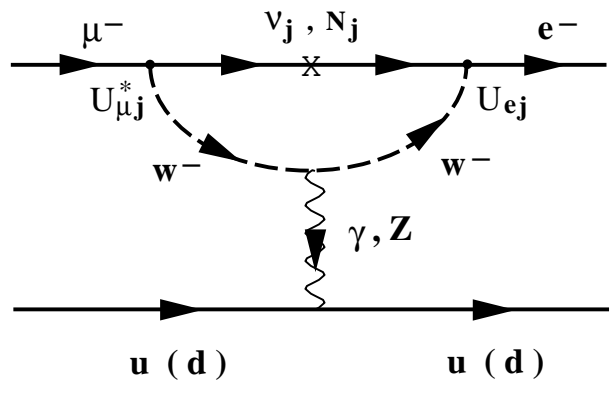
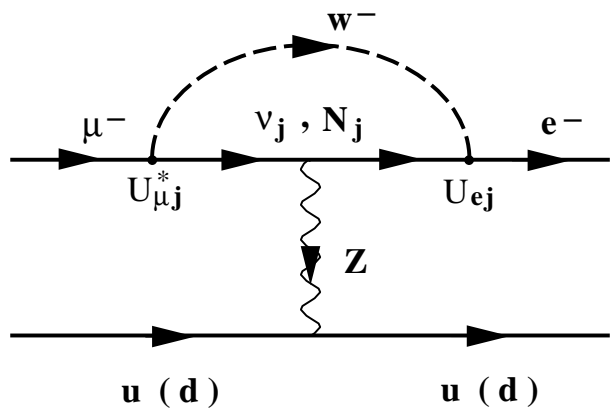
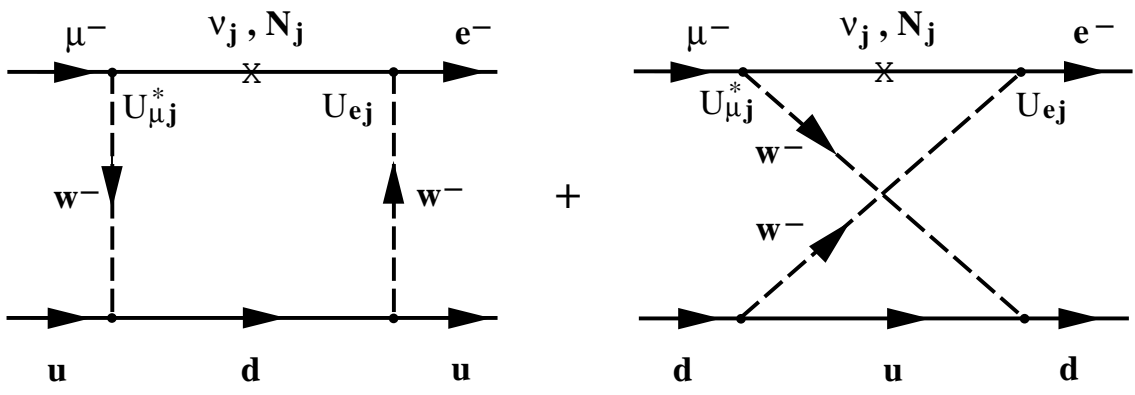
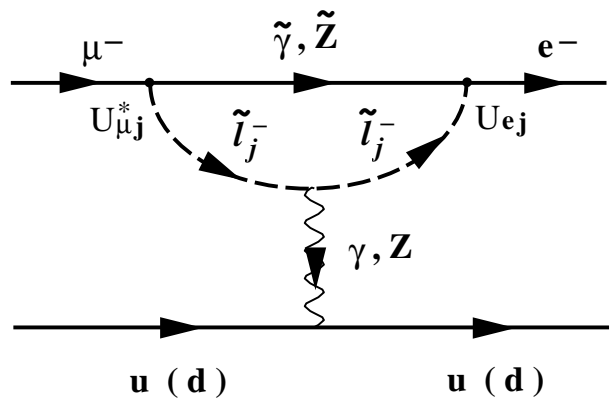
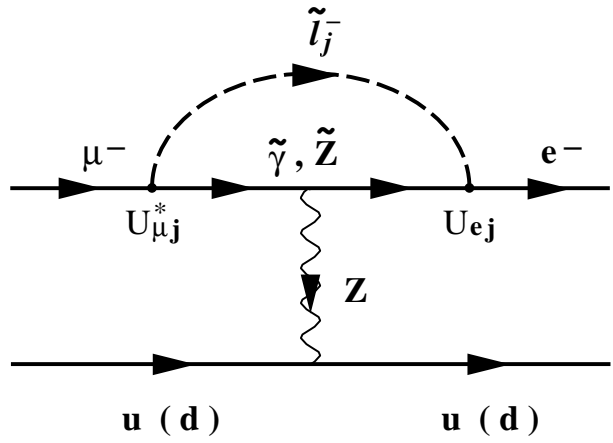
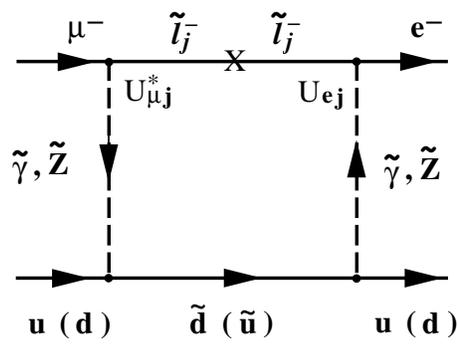
\end{document}